\documentclass[a4paper,11pt]{article}
\usepackage{pos}

\title{Perspectives for particle identification in ALICE using silicon-based timing detectors}
\ShortTitle{Particle identification in ALICE using silicon timing detectors}

\manuallySeparateAuthors
\author*[a]{Roberto Preghenella}
\author{on behalf of the ALICE Collaboration}
\affiliation[a]{Istituto Nazionale di Fisica Nucleare, Sezione di Bologna}

\emailAdd{preghenella@bo.infn.it}

\abstract{Precision timing has alway been a prominent topic in the instrumentation for high-energy physics experiments. In this document I discuss what the perspectives are for the use of silicon sensors for a time-of-flight detector in a next-generation heavy-ion experiment at the LHC. A brief overview of the experiment and a preliminary assessment of the expected particle-identification performance is also presented.}

\FullConference{%
  The Eighth Annual Conference on Large Hadron Collider Physics-LHCP2020 \\
  25-30 May, 2020\\
  online}

\begin{document}
\maketitle

\section{Introduction}

The use of precision-timing technology in high-energy physics (HEP) experiments is rather diverse and covers, for instance, applications in trigger detectors as well as applications in particle-identification, most notably with time-of-flight detectors. In the last two decades resistive plate chambers (RPCs) have been widely used for large-area detectors in place of scintillators, thanks also to the technological advances achieved with multigap-RPC detectors in bringing excellent time resolution performance~\cite{CerronZeballos:1995iy}.

Recently silicon sensors have become very popular for timing in HEP, with a clear application at the high-luminosity LHC for the rejection of the large in-bunch pileup. It is also noteworthy that there is a rapid progress in the development of silicon timing sensors for general consumers that cover a vast range of applications in the fields of imaging, automotive and development of 3D scanners.

\section{ALICE and the next-generation experiment}

The ALICE experiment~\cite{Aamodt:2008zz,Abelevetal:2014cna} is a dedicated heavy-ion experiment at the LHC. The detector, which is currently being upgraded for operations at the LHC in Run3 and Run4, is designed explicitly for heavy-ion physics in order to cope with very high multiplicities and track particles down to very low $p_{\rm T}$. Particle identification (PID) is an essential asset of ALICE, which uses multiple PID techniques and is equipped with a large time-of-flight (TOF) detector in the central barrel for hadron identification up to $p_{\rm T} \sim 5$ GeV/$c$~\cite{Carnesecchi:2018oss}.

The ALICE experiment is being upgraded for operation in LHC Run3 and Run4 with instantaneous luminosities a factor 10-100 larger than then ones recorded so far. Eventually, such luminosities would saturate the maximum interaction rate achievable to operate the current ALICE experiment, whose tracking is based on a large time-projection chamber (TPC) detector. In fact, if another step ahead is to be done in increasing the luminosity capabilities of the experiment, the use of TPC tracking should be abandoned in favour of new technologies. For this reason, a completely new detector to be installed during the LHC long shutdown 4 (LS4) has been proposed~\cite{Adamova:2019vkf} for the future heavy-ion experimental programme at the LHC. The new detector would enable a very rich heavy-ion physics programme in the 2030's, with measurements ranging from electromagnetic probes at ultra-low transverse momentum to precision physics in the charm and beauty sector. As an example, the ability to measure the production of electrons down to $p_{\rm T}$ scales of a few tens of MeV/$c$ would allow one to probe the electromagnetic radiation emitted by the QGP via low-mass dileptons and test theoretical predictions in a region of the phase space where most of the radiation is emitted, which is currently beyond reach. Another example of the physics potential involves the measurement of the production of extremely rare probes, like multiply heavy flavoured (MHF) baryons and mesons. The observation and precise quantification of the production of such exotic hadrons provides a direct window on hadron formation, where spectacular effects are expected for the yields of MHF baryons with respect to the total produced charm quarks, and would represent a tremendous jump for the study of the properties of deconfined QCD matter.

\begin{figure}[t]
  \centering
  \includegraphics[width=0.64\textwidth]{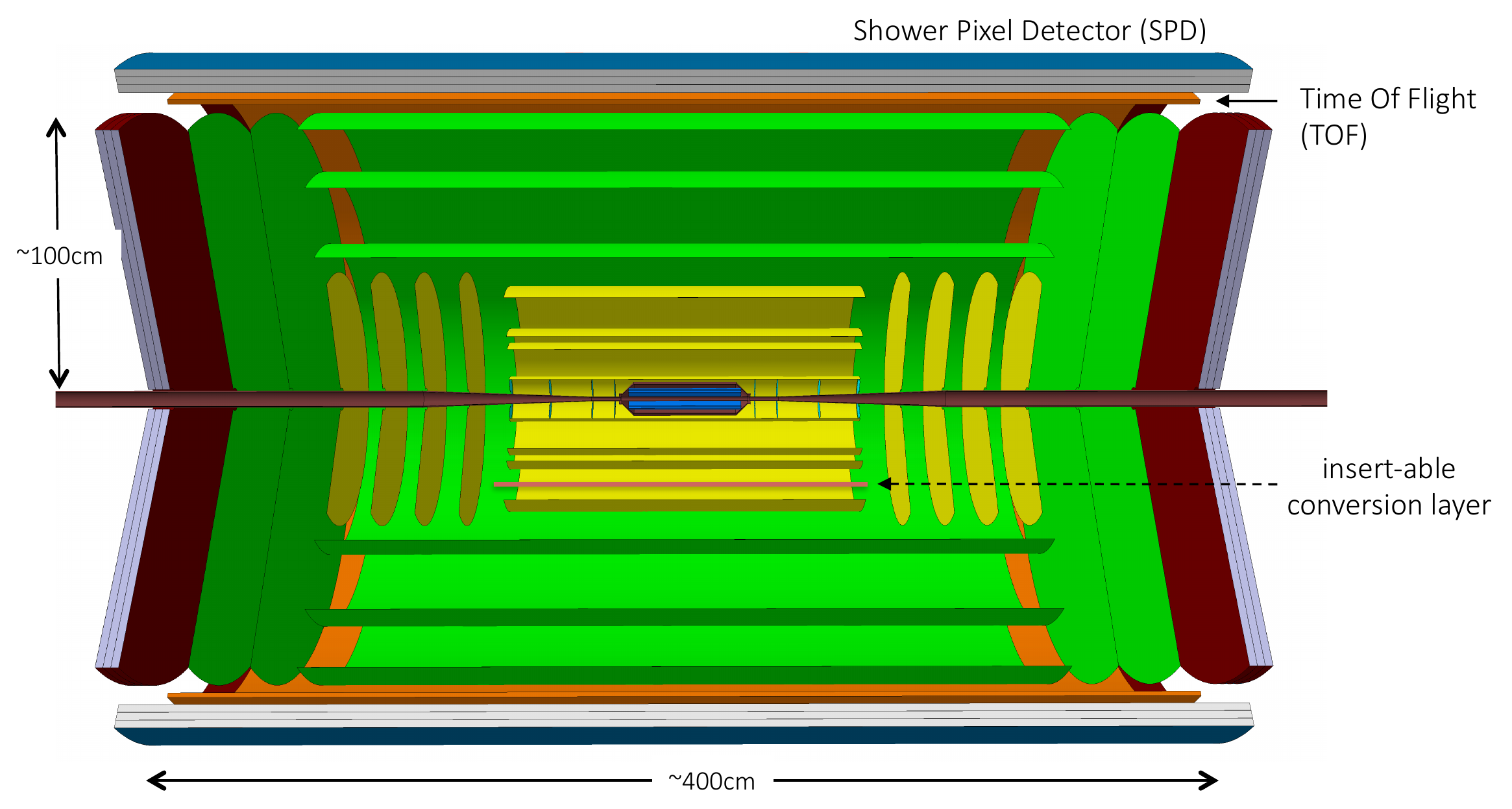}
  \includegraphics[width=0.34\textwidth]{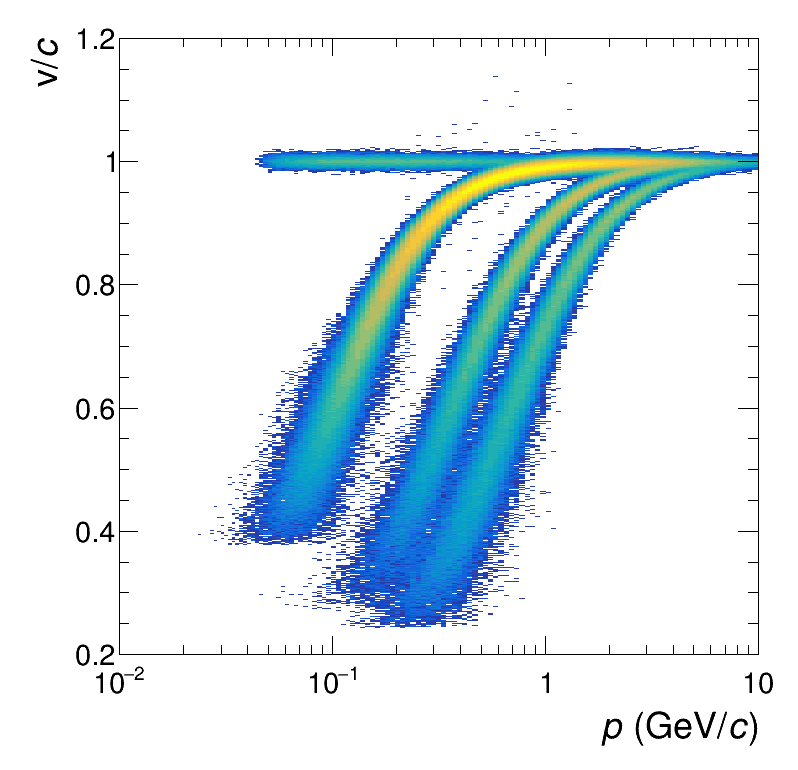}
  \caption{(left) Detector concept for the next-generation heavy-ion experiment at the LHC. (right) Simulated TOF response: measured particle velocity as a function of the reconstructed momentum with the time-of-flight layer.}
  \label{figure1}
\end{figure}

Figure~\ref{figure1} (left) shows the proposed detector concept for a next-generation heavy-ion experiment. It consists of approximately 10 layers of silicon detectors for tracking in the central barrel and in the endcaps and includes a TOF layer also made of silicon detectors to provide hadron identification over a large $p_{\rm T}$ range and electron identification at low $p_{\rm T}$. The detector layout is very compact (radius $\sim$ 1.2 m, length $\sim$ 4 m) and features very low-mass tracking layers ($\sim$ 0.05 $X/X_{0}$ for the innermost vertexing layers, $\sim$ 0.5 $X/X_{0}$ for the tracking layers). It will be a large acceptance ($|\eta|$ < 4, full azimuth) detector with a very-low $p_{\rm T}$ cutoff capable of recording very high nucleon-nucleon luminosities ($\langle \mathcal{L_{\rm NN}} \rangle \sim 10^{34}$ cm$^{-2}$ s$^{-1}$). The performan of the proposed detector features very high resolution for position (< $3 \mu$m for the innermost vertexing layers, $\sim$ $5 \mu$m for the tracking layers) and for time-of-flight ($\sim$ 20 ps) measurements.

\section{Particle-identification with silicon timing detectors}

The proposed next-generation heavy-ion experiment at the LHC will be equipped with a 20 ps timing layer located at approximately 100 cm from the beam line. There are currently multiple technologies under development for high-performance timing sensors. LGAD devices, which have been developed for high-energy physics and are going to be part of the LS3 upgrades of ATLAS and CMS, have proven to be able to provide 20-30 ps resolution also with high radiation loads up to 10$^{13-14}$ 1 MeV neutron equivalent~\cite{OteroUgobono}. SPAD sensors are another possibility for precision timing and come with the advantage of being produced in industrialised facilities and therefore with the possibility to have developments for an integrated monolithic sensor technology~\cite{Veerappan}. Another example of an attractive solution for timing is the monolithic pixel prototype developed by the University of Geneva that reached a time resolution of 50 ps without a gain layer~\cite{Iacobucci:2019rpt} and that promises improved performance with the addition of gain.

One important point to be kept in mind is the lower radiation load expected on the timing layer of the new experiment compared to the levels reached by ATLAS and CMS, a fact that allows one to explore new technology options and therefore perform a complementary R\&D program.

\begin{figure}[t]
  \centering
  \includegraphics[width=\textwidth]{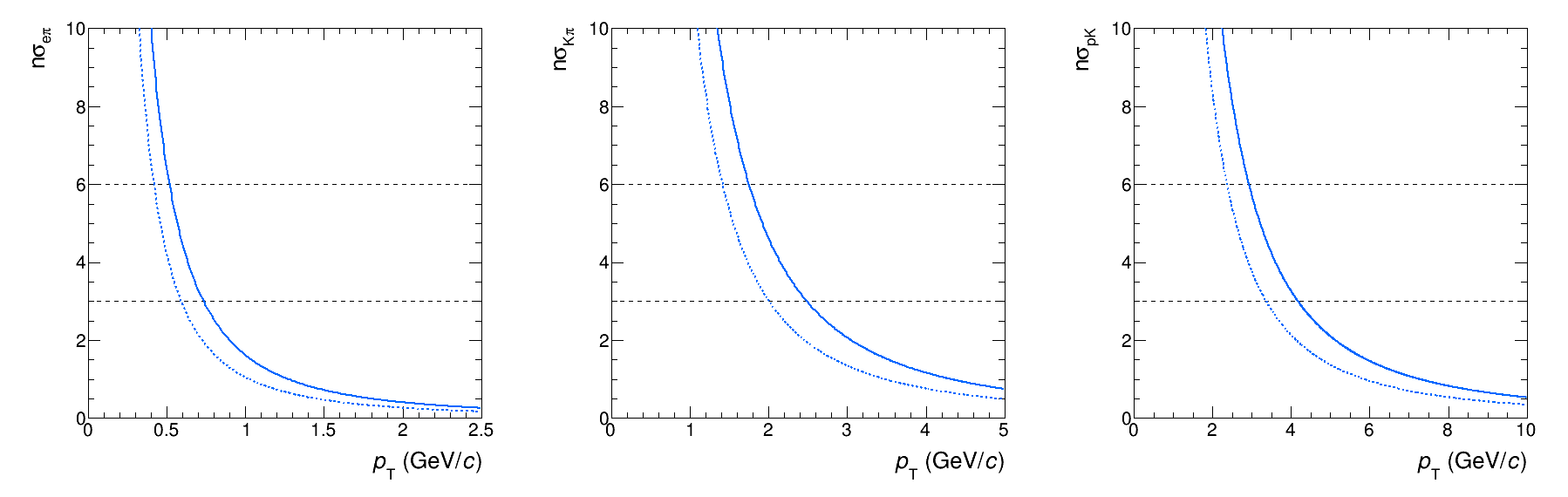}
  \caption{Analytical results of the particle-identification power of the time-of-flight layer in terms of e/$\pi$, K/$\pi$ and p/K number-of-sigma separation at $\eta$ = 0 (solid) and $\eta$ = 1 (dashed).}
  \label{figure2}
\end{figure}

Initial studies on the expected performance of the time-of-flight layer have been presented at this conference. Their goal is to provide an initial guidance for the future optimisations of the detector towards the achievement of the requirements for physics, for example in the identification of electrons at low transverse momentum that are key for the study of the thermal radiation from the QGP with low-mass dileptons.

Analytical estimates of the performance can be obtained and one can express the separation power in terms of the number of standard deviations (n$\sigma$) for a 20 ps timing layer located at 100 cm from the beam line, as shown in Figure~\ref{figure2}. At midrapidity ($\eta$ = 0) the system can provide a better than 3$\sigma$ $e/\pi$ separation up to about $p_{\rm T}$ = 750 MeV/$c$ and up to about $p_{\rm T}$ = 2.5 GeV/$c$ for K/$\pi$ separation. The performance estimates here are based on a simplified simulation which only takes into account the configuration and layout of the tracker detector, including material effects, in a fast-simulation Monte Carlo approach. A more refined estimate of the performance requires a full simulation and reconstruction and is not available yet. Overall, despite some of the effects (e.g. bad hit association) need to be added, these studies provide a reliable estimate and valuable guidance on how the full system would perform. Figure~\ref{figure1} (right) shows the response of the time-of-flight layer when coupled with a possible configuration of the tracker embedded in a 0.2 T solenoidal magnetic field. 

\section{Conclusions}

There are several interesting ongoing developments for timing with silicon sensors and promising technologies, which can be further optimised for particle identification with the time-of-flight technique. A next-generation experiment has been proposed for heavy-ion research at the LHC in the 2030s and that will be equipped with a high-performance time-of-flight layer. The different requirements in terms of radiation load with respect to ATLAS and CMS open the possibility to explore, research and develop new technologies for the timing sensors that the experiment can adopt.

%\begin{figure}[t]
%  \centering
%  \includegraphics[width=0.8\textwidth]{PIDeffCont.png}
%\end{figure}

\end{document}